\documentclass{PoS}

\title{The Calibration System of the HAWC Gamma-Ray Observatory}
\ShortTitle{HAWC Calibration System}

\author{H. A. Ayala Solares$^a$, M. Gerhardt$^a$, C. M. Hui$^a$, R. J. Lauer$^b$, Z. Ren$^b$, F. Salesa Greus$^c$, and H. Zhou$^{a}$ for the HAWC Collaboration$^d$\\
\llap{$^a$}Department of Physics, Michigan Technological University, 1400 Townsend Drive, Houghton, MI, USA\\
\llap{$^b$}Deptartment of Physics and Astronomy, University of New Mexico, 1919 Lomas Blvd. NE, Albuquerque, NM, USA\\
\llap{$^c$}Department of Physics, Pennsylvania State University, 16802 University Park, PA, USA\\
\llap{$^d$}For a complete author list, see \href{http://www.hawc-observatory.org/collaboration/icrc2015.php}{www.hawc-observatory.org/collaboration/icrc2015.php}.\\ 
E-mail: \email{hayalaso@mtu.edu}, \email{mjgerhar@mtu.edu}, \email{cmhui@mtu.edu}, \email{rlauer@phys.unm.edu}, \email{zxren@unm.edu}, \email{sagreus@psu.edu}, \email{hzhou1@mtu.edu}}

\abstract{The HAWC collaboration has recently completed the construction of a gamma-ray observatory at an altitude of 4100 meters on the slope of the Sierra Negra volcano in the state of Puebla, Mexico. In order to achieve an optimal angular resolution, energy reconstruction, and cosmic-ray background suppression for the air showers observed by HAWC, it is crucial to obtain good timing and charge calibrations of the photosensors in the detector. The HAWC calibration is based on a laser system which is able to deliver short light pulses to all the tanks in the array. The light intensity can range over 7 orders of magnitude, broad enough to cover all the dynamic range of the PMT readout electronics. In this contribution we will present the HAWC calibration system, together with the methods used to calibrate the detector.}

\FullConference{The 34th International Cosmic Ray Conference,\\
		30 July- 6 August, 2015\\
		The Hague, The Netherlands}

\begin{document}

\section{Introduction}
\label{sec:intro}

Inaugurated on March 19$^{th}$ 2015, the HAWC Gamma-Ray Observatory is presently fully operational with 300 Water Cherenkov Detectors (WCDs) covering an area of 22,000 m$^{2}$ on the northern slope of the Sierra Negra volcano (4100 m.a.s.l.) in the state of Puebla, Mexico.

The performance of the detector, and therefore the sensitivity, depends critically on an accurate calibration of the PMTs in the detectors. The HAWC collaboration has developed a calibration system that consists of a pulsed laser able to illuminate all the WCDs through a network of optical fibers. In the following sections we describe the hardware and the charge and timing calibration methods presently used in HAWC.

\section{Hardware Description}
\label{sec:hardware}
A 532\,nm wavelength laser is used to calibrate the 1200 PMTs in the HAWC array. The laser produces pulses pulses with energies of 45\,$\mu$J and widths of 300\,ps with a frequency of 200\,Hz.  Neutral density filters are used to provide different light levels with over 7 orders of magnitude dynamic range of attenuation.  This allows the laser setup to calibrate PMTs between 1\,photo-electron (PE) to over thousands of\,PEs.  The laser intensity is measured with a silicon energy probe and a radiometer at different parts of the setup for light output measurement and monitoring.  The laser light is fanned out to each individual WCD using a network of optical splitters, fiber optic switches, and fiber optic cables.  Inside each WCD, a diffuser is used to distribute the light to the 4 PMTs within.  

\begin{figure}[h]
\centering
\includegraphics[width=0.9\textwidth]{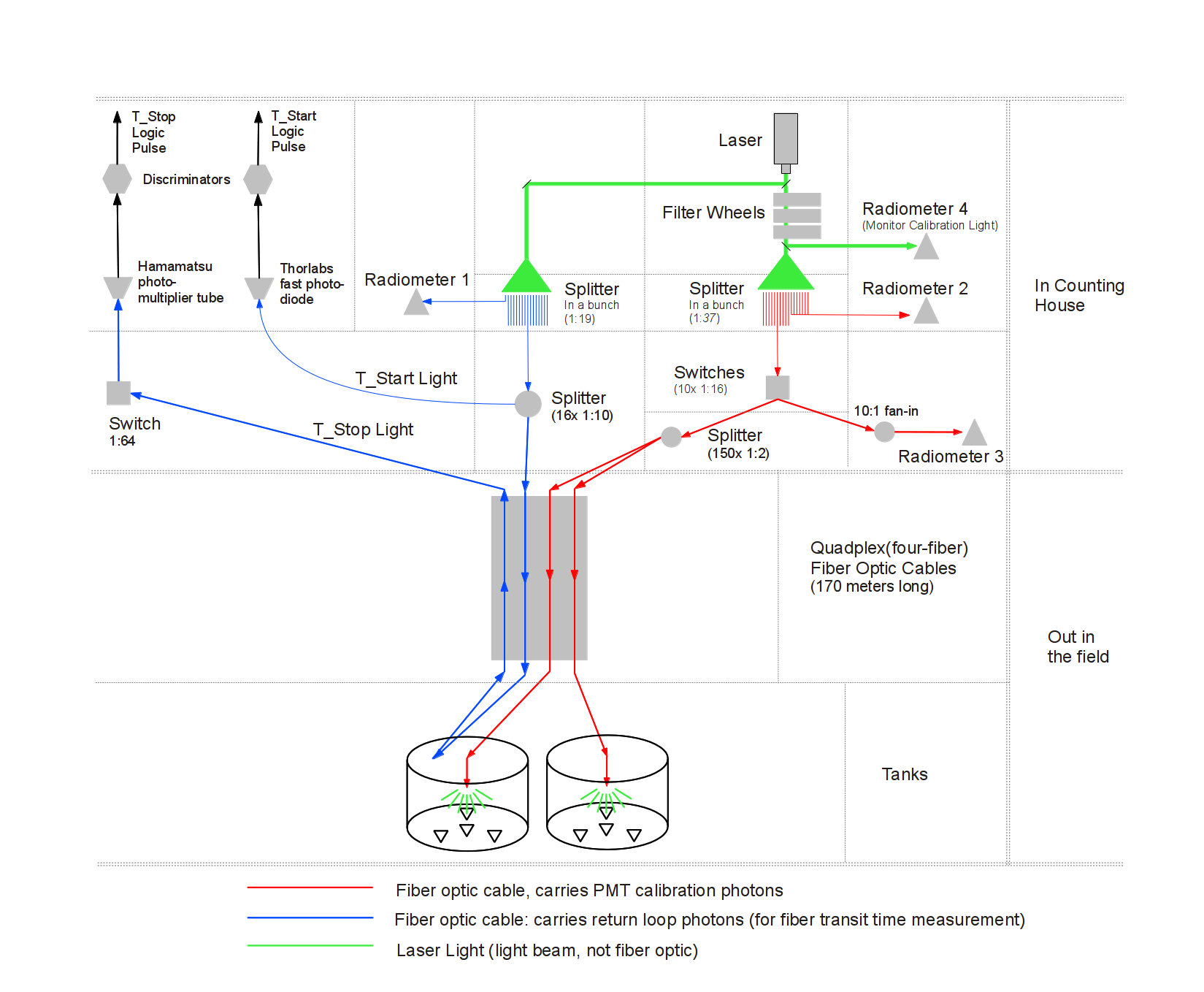}
\caption{Schematic of the calibration setup for HAWC.}
\label{fig:CalSetup}
\end{figure}

A pulser is used to trigger the laser and the data acquisition system (DAQ) of HAWC.  Together with the laser light split before the neutral density filters, a start signal is sent to the DAQ to flag the event as calibration.  Figure \ref{fig:CalSetup} shows a schematic of the calibration system setup.

\section{Charge Calibration}
\label{sec:charge}

An air shower crossing the detectors produces signals in the PMTs of the array. These signals are digitized if they are large enough to cross the discriminator threshold. The time difference between the trailing and leading edges of the signal is called Time-over-Threshold (ToT). The ToT is related to the amplitude and therefore to the charge of the pulse in the PMT.
The ultimate goal of the charge calibration is to translate the ToT units recorded by the electronics into charge units, in PEs.
Calibration runs can deliver different amounts of light to the PMTs using several filter wheel (FW) combinations, which allow us to scan the whole ToT dynamic range.

\subsection{Occupancy method and PE determination}

When laser light is sent to a PMT, the collected charge can be determined using the occupancy method, which was introduced in ref.~\cite{IMB}. 
The occupancy is defined as the fraction of events emitted by the laser which are observed in the PMT. 
When the discriminator threshold is set low enough to produce a signal if at least one PE arrives to the photocathode, the occupancy is the probability of observing more than 0 PEs:

\begin{equation}
\eta = \frac{\rm{signal}}{\rm{trigger}}=P(n>0,<n_{PE}>)=1-P(n=0,<n_{PE}>)=1-e^{-<n_{PE}>}
\label{eq: npe-Occ-eq}
\end{equation}

\noindent where $\eta$ is the occupancy and $<n_{\rm{PE}}>$ is the mean number of PEs observed.

\begin{figure}[htpb]
\begin{center}
\includegraphics[width=0.65\linewidth, angle=0]{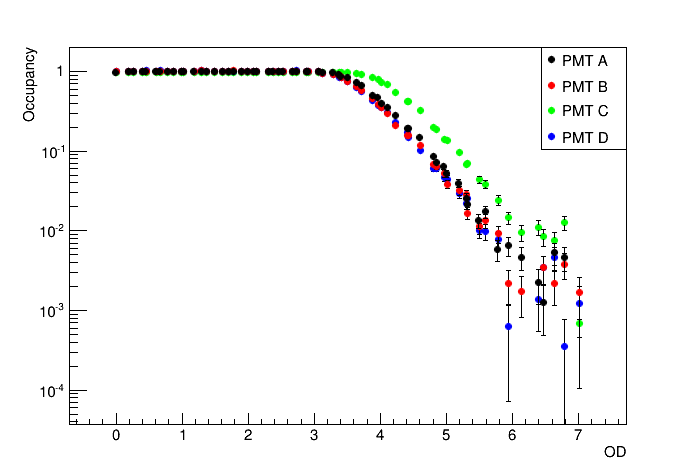}
\caption{Occupancy as a function of the OD (attenuation of the laser beam, see text.) covering 7 orders of magnitude in ODs and 3 orders of magnitude in occupancy. Each point represents one particular FW setting, i.e., one particular laser light intensity.}
\label{fig: Occ_plot}
\end{center}
\end{figure}

Figure~\ref{fig: Occ_plot} shows the occupancy from the calibration data for one WCD taken in April 2015. The plot shows the occupancy as a function of the optical depth (OD) of each FW combination (63 in total). The OD is directly related to the transmittance (fraction of light delivered, being 1 the maximum) via the formula $T=10^{-OD}$. 
Each WCD in HAWC has three 8-inch PMTs in a triangular layout and a 10-inch PMT located at the center of this triangle. As it can be seen in Figure~\ref{fig: Occ_plot}, the central PMT (green dots) has larger occupancy. This is due to the larger photocathode, higher efficiency, and its location, closer to the diffuser. 

Measuring the occupancy, we can directly obtain $<n_{\rm{PE}}>$ from Eq.~\ref{eq: npe-Occ-eq}:

\begin{equation}
<n_{PE}>=-ln(1-\eta)
\label{eq: npe-eq}
\end{equation}

Getting $<n_{\rm{PE}}>$ from the occupancy works fine at low intensity levels, i.e., below 2 PEs. Unfortunately, this method cannot be used at high intensities because the error increases with the exponential of $<n_{\rm{PE}}>$.
However, we can assume that $<n_{\rm{PE}}>$ detected by a PMT is proportional to the light intensity at the photocathode, so we can use a linear relation between the intensity and the number of PEs, and assume this is also true for intensities above 2PEs.

The parameters of the linear relation are obtained fitting the $<n_{\rm{PE}}>$ that are below 2PEs, i.e., obtained from the occupancy method to:

\begin{equation}
<n_{PE}> =A(I/I_{ref})+B
\label{eq: trans-I}
\end{equation}

\noindent The values of the intensity (I) come from the energy measurements of the radiometers (see sketch in Fig.~\ref{fig:CalSetup}). I$_{ref}$ is the reference value for the intensity. We can pick for instance the maximum intensity in the calibration run ($\sim 10^{-8}$ J) as the reference. This choice permits us to identify the parameter $A$ as the maximum number of PEs for which a particular PMT is calibrated. 

\subsection{ToT-PE conversion}

After the determination of the $<n_{\rm{PE}}>$ at a given laser intensity, the next step is to simulate the distribution of the actual number of PEs to mimic the PMT response. In order to achieve this we smear the discrete values coming from the Poisson statistics using a Gaussian of width $0.35 \sqrt{<n_{\rm{PE}}>}$. The 35\% factor comes from the energy resolution of the PMTs.  

\begin{figure}[htpb]
\begin{center}
\includegraphics[width=0.65\linewidth, angle=0]{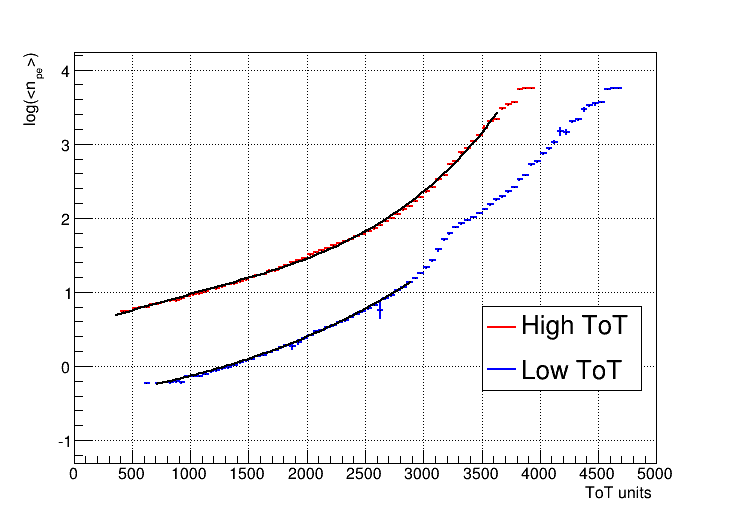}
\caption{Plot showing the log($<n_{PE}>$) as a function of the ToT counts. The red dots are the High ToT and the blue ones the Low ToT.}
\label{fig: ToT_plot}
\end{center}
\end{figure}

Then the simulated $<n_{\rm{PE}}>$ distribution is compared to the actual ToT distribution recorded by the DAQ for each intensity (FW setting). Matching the percentiles in 10\% steps from 10\% to 90\% for both distributions we obtain for each PMT a figure like Fig.~\ref{fig: ToT_plot}. This figure shows the profile plot of the base-10 logarithm of the percentiles of the simulated charge distribution, with respect to the percentiles of the actual ToT distribution. The two sets of points correspond to the so-called High and Low thresholds, which are the two discriminator thresholds used in the HAWC DAQ. This plot was made using 63 FW combinations.   

Finally, the points are fitted to a broken power-law with 6 parameters. This fitted function is the one used by the reconstruction algorithms to convert the ToT into PEs for each signal in that PMT.

\subsection{Charge calibration performance}

\begin{figure}[htpb]
\begin{center}
\includegraphics[width=0.65\linewidth, angle=0]{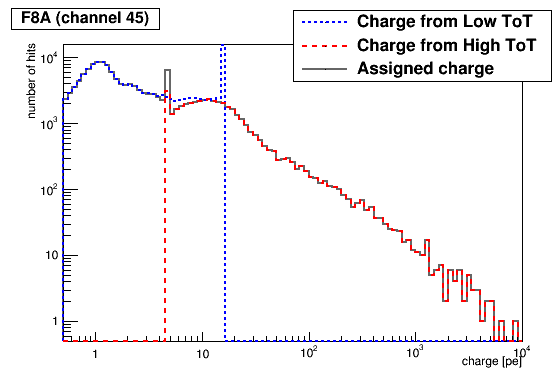}
\caption{Charge rate distribution in log-log scale of an example PMT. The blue histogram represent the low ToT signal and the red histogram represent the high ToT signal. High ToT calibration is used for most signals in the overlap region (in gray).}
\label{fig: charge_dis}
\end{center}
\end{figure}

When the calibration of ToT to PE is applied to triggered data we obtain the charge distribution for all the PMTs in HAWC. In Figure ~\ref{fig: charge_dis} we show an example for a particular PMT.
The charge rate distribution has features such as a single photo-electron (SPE) peak associated with a typical width (sigma). The physical reason of the SPE peak is the high probability of a particle generating only one photo-electron. The width of the SPE is related to the uncertainty of the charge measurement. The dip seen in the assigned charge comes from using two thresholds (Low ToT and High ToT). All these features are used to understand the performance of the PMTs. The charge rate distributions are used to point out potential bad channels. Moreover, some efforts are on-going to use the fatures of the rate distribution to point out specific hardware issues.

The Low ToT range of the charge distribution can be fitted with a double Gaussian model: 

\begin{equation}
f(x) = p[0]\exp(-\frac{(x-p[1])^2}{2{p[2]^2}}) + p[3]\exp(-\frac{(x-2p[1])^2}{4{p[2]^2}})
\label{eq: npe-Occ-eq}
\end{equation}

\noindent where $p[0]$ is the normalization of SPE Gaussian, $p[1]$ is the SPE value and $p[2]$ is the width of SPE Gaussian. 

The Figure ~\ref{fig: peak_dis} below shows the distribution of the parameters. Note that the blue histogram represents the 8 inch PMTs and the magenta histrogram represents the 10 inch PMTs. The green and red lines represent the 4 sigma cuts applied to those distributions. The SPE distribution shows that the mean value of both 8 inch and 10 inch PMTs are around 1, which provides a check for the accuracy of the charge calibration. The mean value of the sigma distribution is around 35\%, which is also consistent with the assumed PMT energy resolution. The outliers PMTs can easily be identified relative to the 4 sigma intervals indicated by red and green lines. An automated way of checking bad PMTs (based on charge distribution) has been developed.

\begin{figure}[htpb]
\begin{center}
\begin{tabular}{c c}
\hspace{-0.6 cm}
\includegraphics[width=0.53\linewidth, angle=0]{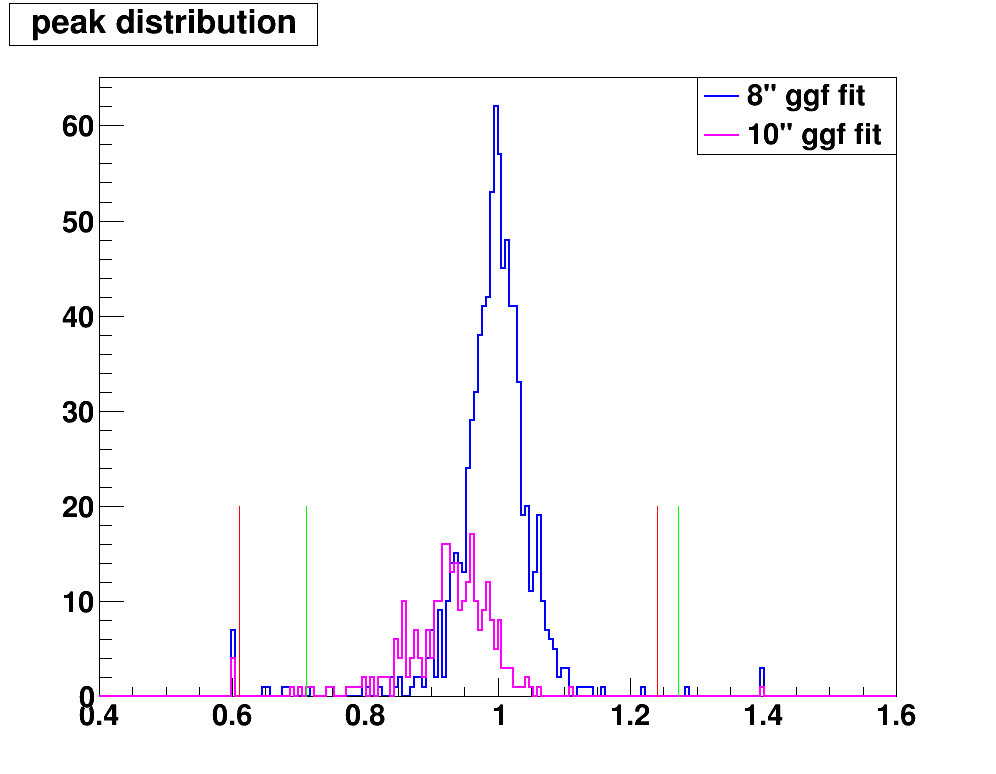}
\hspace{0.0 cm}
\includegraphics[width=0.53\linewidth, angle=0]{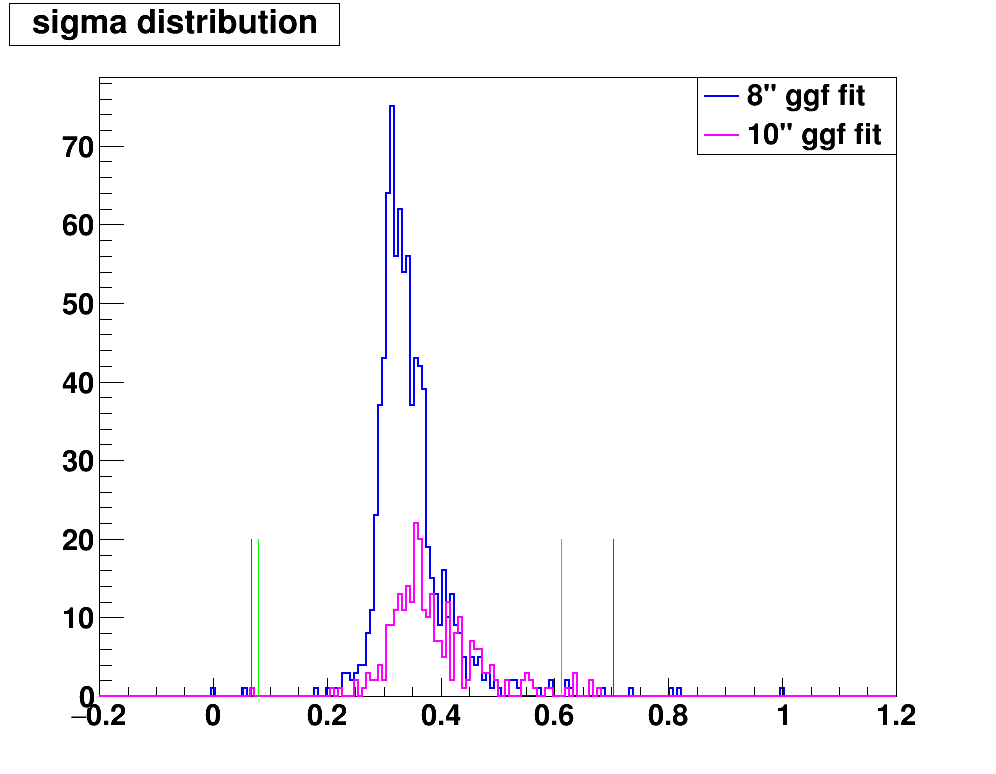}
\end{tabular}
\caption{The left figure is the SPE distribution of all PMTs from an example sub run. The right figure is the sigma distribution of all PMTs.}
\label{fig: peak_dis}
\end{center}
\end{figure}

\section{Timing Calibration}
\label{sec:timing}
Accurate ($\sim$1 ns) timing calibration is required for the reconstruction of the arrival direction of the primary particles that produce the extensive air showers. This does not require absolute timing but does require accurate relative timing between PMT channels. Timing calibration is divided into three parts: correcting for the different light paths, for the PMT slewing time, and for the time residual with respect to the shower front.
The timing to account for light paths contains two main parts: the light transit time of the optical fibers, and the geometry of the PMT setup within the WCD.  

\subsection{Slewing} 
\begin{figure}[h]
\centering
\includegraphics[width=0.65\textwidth]{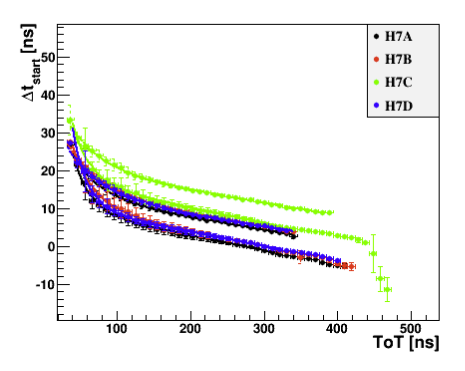}
\caption{Slewing time (in ns) as a function of the ToT (in ns) for the PMTs in tank H7. The two set of point for each PMT represent the two thresholds. H7C is the central PMT.}
\label{fig:slewplot}
\end{figure}

As we mentioned before, the HAWC electronics uses two thresholds to sample small (Low) and large (High) PMT pulses. When a pulse crosses a threshold it is given a time stamp by the time-to-digital converter (TDC) with a precision of 0.1\,ns.~\cite{ICRC} Slewing time is caused by the dependence of the rise time with the pulse height. The larger the pulse, the earlier it crosses the first threshold and the larger the ToT. The time between PMT hit time and the first threshold crossing, or slewing time, is given by equation~\ref{eq: Slewing} for a given ToT. Figure \ref{fig:slewplot} shows a typical slewing curves for the 4 PMTs in a WCD.
\begin{equation}
\textrm{Slewing Time} = e^{\frac{-ToT-p0}{p1}}-e^{\frac{ToT-p2}{p3}}+p4-p5*ToT
\label{eq: Slewing}
\end{equation}

\subsection{Time Residuals}
\begin{figure}[h]
\centering
\includegraphics[width=0.65\textwidth]{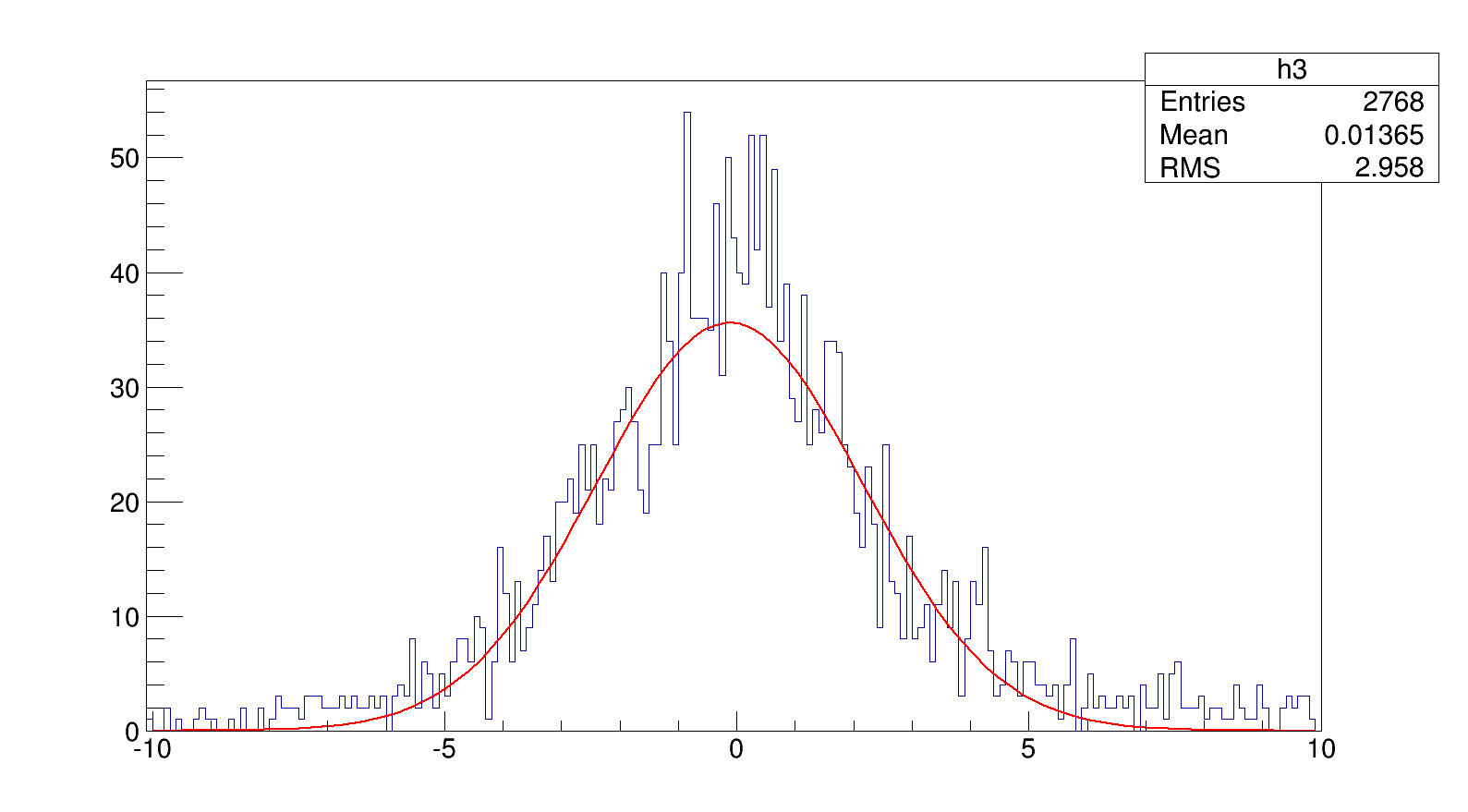}
\caption{\textbf{Time residual fit for H7C}}
\label{fig:trplot}
\end{figure}

The time residual is an offset between the PMT time and the fitted air shower front expected time.
This is calculated after reconstructing the air shower core position and direction from 15000 airshower events with nHit>500. Shown in Figure \ref{fig:trplot} is a histogram of the time residual for a single PMT, the peak value from the Gaussian fit indicates the PMT is either early or late compared to the shower plane. This PMT is then assigned a time pedestal value such that in the next iteration the time residual will be closer to zero. This process is repeated 3 times, to reduce the bias of the initial value of the time residuals, to achieve a 0.1\,ns accuracy in timing. After the pedestal is produced, alignment with zenith, directly overhead, is checked by taking a sample of reconstructed data to determine the direction of the reconstructed zenith angle for the sample. An additional time residual is produced, for each PMT based upon its location in the array and the reconstructed zenith found from the sample, which is added to the existing time pedestal to make the detector point overhead. 

\section{Conclusions}
\label{sec:conclusions}
In these proceedings we have explained briefly the laser system used in HAWC for the PMT calibration. This system is currently able to calibrate the full HAWC array. 
We have explained the charge calibration, based in the occupancy method, which allow us to calibrate all the PMTs from fractions to thousands of PEs. We have also shown that the laser system allows the calibration of the PMTs to better than 1ns.
Checking the charge rate distribution and the time residuals of the PMTs we have verified that the system performs as expected.

\section*{Acknowledgments}
\footnotesize{
We acknowledge the support from: the US National Science Foundation (NSF);
the US Department of Energy Office of High-Energy Physics;
the Laboratory Directed Research and Development (LDRD) program of
Los Alamos National Laboratory; Consejo Nacional de Ciencia y Tecnolog\'{\i}a (CONACyT),
Mexico (grants 260378, 55155, 105666, 122331, 132197, 167281, 167733);
Red de F\'{\i}sica de Altas Energ\'{\i}as, Mexico;
DGAPA-UNAM (grants IG100414-3, IN108713,  IN121309, IN115409, IN111315);
VIEP-BUAP (grant 161-EXC-2011);
the University of Wisconsin Alumni Research Foundation;
the Institute of Geophysics, Planetary Physics, and Signatures at Los Alamos National Laboratory;
the Luc Binette Foundation UNAM Postdoctoral Fellowship program.
}

\end{document}